\documentclass[english,aps, superscriptaddress, floatfix, 12pt]{revtex4}
\usepackage[T1]{fontenc}
\usepackage[latin1]{inputenc}
\usepackage{graphicx}
\usepackage{amssymb}
\usepackage{epsfig}
\usepackage{amsmath}
\usepackage{psfrag}
\usepackage[active]{srcltx}
\usepackage{babel}
\def\be{\begin{equation}}
\def\ee{\end{equation}}
\def\bea{\begin{eqnarray}}
\def\eea{\end{eqnarray}}
\makeatother   
\begin{document}
\title{Quark condensate for various heavy flavors}
\author{Dmitri Antonov and Jos\'e Em\'ilio F.T. Ribeiro\\
{\it Departamento de F\'isica and Centro de F\'isica das Interac\c{c}\~oes Fundamentais,}\\ 
{\it Instituto Superior T\'ecnico, UT Lisboa,
Av. Rovisco Pais, 1049-001 Lisboa, Portugal}}

\noaffiliation

\begin{abstract}
The quark condensate is calculated within the world-line effective-action formalism, by using for the Wilson loop an
ansatz provided by the stochastic vacuum model. Starting with the relation between 
the quark and the gluon condensates in the heavy-quark limit, we diminish the current quark mass down to 
the value of the inverse vacuum correlation length, finding in this way a 64\%-decrease in the 
absolute value of the quark condensate. In particular, we find that the conventional formula for the heavy-quark 
condensate cannot be applied to the $c$-quark, and that the corrections to this formula can reach 
23\% even in the case of the $b$-quark. 
We also demonstrate that, for an exponential parametrization of the 
two-point correlation function of gluonic field strengths, 
the quark condensate does not depend on the non-confining non-perturbative interactions of the stochastic background Yang--Mills fields. 
\end{abstract}

\maketitle

\section{Introduction}
As it is well known, the so-called chiral SU$_L(N_{\rm f})\times$SU$_R(N_{\rm f})$ 
symmetry of the classical action of 
QCD with $N_{\rm f}$ massless quark flavors is spontaneously broken at the quantum level, with the order parameter
for this symmetry breaking being the quark condensate $\langle\bar\psi\psi\rangle$.
Together with confinement, which is characterized by the gluon condensate $\langle (gF_{\mu\nu}^a)^2\rangle$, chiral-symmetry breaking is one of the two most important non-perturbative phenomena in QCD. 
A natural question can be posed as whether these phenomena are interrelated or not. 
An affirmative answer to this question would imply proportionality between the quark and the gluon condensates, 
which was indeed found in Ref.~\cite{ne}. The corresponding relation reads
\be
\label{kj}
\langle\bar\psi\psi\rangle\propto
-T_g{\,}
\langle (gF_{\mu\nu}^a)^2\rangle,
\ee 
where $T_g$ is the vacuum correlation length, at which the 
two-point gauge-invariant correlation function of gluonic field strengths exponentially falls off. 
Equation~(\ref{kj}) stems from the integration in the QCD partition function 
over soft gluonic fields in the leading, Gaussian, approximation. Within this approximation, the kernel of the four-quark interaction is defined by the two-point field-strength correlation function with the amplitude $\langle (gF_{\mu\nu}^a)^2\rangle$ and the correlation length $T_g$.

Alternatively, if one first integrates in the QCD partition function over the quark fields, one arrives at a
gauge-invariant effective action, where the gluonic degrees of freedom are represented in the form of Wilson loops and their correlation functions~\cite{chr}. An advantageous feature of this approach is that,
owing to the color-neutrality of Wilson loops, the calculation of the effective action becomes reduced to the calculation of the world-line integrals in an {\it Abelian} gauge theory. 
When the dynamical quarks which are integrated out are sufficiently heavy, namely their current mass $M$
is larger than $1/T_g$, the gluonic field inside the 
quark trajectory can be treated as nearly constant. In this heavy-quark limit, the one-loop 
effective action yields the following heavy-quark condensate of a given flavor~\cite{2}:
\be
\label{hee}
\langle\bar\psi\psi\rangle_{\rm SVZ}=-\frac{
\langle (gF_{\mu\nu}^a)^2\rangle}{48\pi^2 M}.
\ee 
This expression coincides with the one known from the SVZ sum rules~\cite{3}.

With $M$ decreasing downwards $1/T_g$, variations of the gauge field inside the quark trajectory produce
corrections to Eq.~(\ref{hee}). The aim of the present paper is the calculation of such corrections. 
They will be obtained by using the approach of Ref.~\cite{1}, which, for the case of a fermion moving 
in an {\it arbitrary} Abelian gauge field, yields a 
closed formula for the effective action with two field strengths. 
Furthermore, it is known that, in addition to the confining interactions of stochastic
gluonic fields, there also exist non-confining non-perturbative interactions of those fields, albeit of a relatively
small strength (cf. Ref.~\cite{ri}). Below, we study the influence of such interactions
on the heavy-quark condensate. For the case of the simplest, purely exponential, two-point correlation function of 
gluonic field strengths, we find the interesting phenomenon of a complete independence 
of the heavy-quark condensate from the non-confining non-perturbative interactions.

The paper is organized as follows. In the next Section, we calculate the 
quark condensate by accounting in the effective action for the confining interactions of stochastic gluonic fields.
In Section~III, we generalize this result to the case 
where non-confining non-perturbative interactions of those fields are taken into account.
Section~IV provides a summary of the results obtained.

\section{Corrections to the heavy-quark condensate}

Integrating over the quark fields in the QCD partition function, one arrives at the following one-loop 
effective action~\cite{chr,2,1}:
$$\langle\Gamma[A_\mu^a]\rangle=2N_{\rm f}\int_0^\infty\frac{ds}{s}{\,}{\rm e}^{-M^2s}\times$$
\begin{equation}
\label{effA}
\times
\int_{P}^{}{\cal D}z_\mu \int_{A}^{}
{\cal D}\psi_\mu{\,} {\rm e}^{-\int_0^s d\tau\left(\frac14\dot z_\mu^2+\frac12\psi_\mu\dot\psi_\mu\right)}
\exp\left[-2\int_0^s d\tau{\,}\psi_\mu\psi_\nu\frac{\delta}{\delta \sigma_{\mu\nu}(z)}\right]
\langle W[z_\mu]\rangle.
\end{equation}
Here, $P$ and $A$ stand, respectively, for the periodic and the antiperiodic boundary conditions, so that
$\int_P^{}\equiv\int_{z_\mu(s)=z_\mu(0)}^{}$, $\int_A^{}\equiv\int_{\psi_\mu(s)=-\psi_\mu(0)}^{}$, and $M$ 
is the current quark mass. Since the quark condensate is always associated with a given 
flavor, we set $N_{\rm f}=1$. The corresponding expression for the quark condensate reads
\be
\label{qq}
\langle\bar\psi\psi\rangle=
-\frac{1}{V}{\,}\frac{\partial}{\partial M}\langle\Gamma[A_\mu^a]\rangle,
\ee 
where $V$ is the Euclidean four-volume occupied by the system, and
\be
\label{av4}
\left<\cdots\right>\equiv\int {\cal D}A_\mu^a{\,} (\cdots){\,} 
{\rm e}^{-\frac14\int_x (F_{\mu\nu}^a)^2}
\ee 
is the average over gluonic fields.
In the heavy-quark limit of $M\gg 1/T_g$, the one-loop approximation becomes exact, leading to Eq.~(\ref{hee}) [cf. Ref.~\cite{2} and the paragraph after Eq.~(\ref{uu}) below].
We notice that Eq.~(\ref{effA}) uses the fact that
the Yang--Mills field-strength tensor $F_{\mu\nu}^a=\partial_\mu A_\nu^a-\partial_\nu A_\mu^a-gf^{abc}A_\mu^b A_\nu^c$, which enters the quark spin term in the world-line action, 
can be recovered by means of 
the area-derivative operator $\frac{\delta}{\delta \sigma_{\mu\nu}}$
acting on the Wilson loop~\cite{mm}. By virtue of this fact, all the 
gauge-field dependence of the effective action becomes encoded in the Wilson loop. The latter is defined by the 
usual formula
$\langle W[z_\mu]\rangle=\left<{\rm tr}{\,}{\cal P}{\,}\exp\left(ig\int_0^s d\tau{\,} 
T^aA_\mu^a\dot z_\mu\right)\right>$,
where $T^a$ is a generator of the SU($N_c$)-group in the fundamental representation, and 
${\cal P}$ denotes the path ordering.

Since $\langle W[z_\mu]\rangle$ is completely determined by the geometric characteristics of 
the contour $C\equiv z_\mu(\tau)$, the calculation of 
the quark condensate becomes an Abelian problem. In this section, we consider the 
confining part of $\langle W(C)\rangle$, deferring the study of the subleading non-perturbative 
non-confining part to the next section. 
Within the stochastic vacuum model~\cite{ds}, 
the corresponding area-dependent part of the Wilson loop reads 
\be
\label{ww}
\langle W(C)\rangle = N_c{\,}\exp\left[-\frac{G}{96N_c}\int_\Sigma d\sigma_{\mu\nu}(x)
\int_\Sigma d\sigma_{\mu\nu}(x'){\,}{\rm e}^{-\mu|x-x'|}\right],~~~ {\rm where}~~~ \mu\equiv\frac{1}{T_g}.
\ee
In this formula, $\Sigma$ is the minimal surface bounded by the contour $C$, and  
$G\equiv\langle (gF_{\mu\nu}^a)^2\rangle$ is the gluon condensate. 
Furthermore, we choose the surface element $d\sigma_{\mu\nu}$
in the form of an oriented, infinitely thin triangle built up of the 
position vector $z_\mu(\tau)$ and the differential element
$dz_\mu=\dot z_\mu d\tau$, namely $d\sigma_{\mu\nu}(z)=\frac12(z_\mu\dot z_\nu-z_\nu\dot z_\mu)d\tau$.
One can then readily check that 
$\int d\sigma_{\mu\nu}(z)\int d\sigma_{\mu\nu}(z')=\left(\int_0^s d\tau{\,}\dot z_\mu
z_\nu\right)^2$, as it should be~\footnote{The latter formula can be proved by rewriting the double surface 
integral as 
$$\int d\sigma_{\mu\nu}(x)\int d\sigma_{\mu\nu}(x')=-\frac12\int d\sigma_{\mu\nu}(x)\int d\sigma_{\mu\rho}(x'){\,}
\partial_\nu^x\partial_\rho^{x'}{\,}(x-x')^2,$$
applying the Stokes' theorem, which leads to
$$\int d\sigma_{\mu\nu}(x)\int d\sigma_{\mu\nu}(x')=-\frac12\oint dz_\mu \oint dz_\mu'(z-z')^2,$$
and noticing that only the $(zz')$-term in $(z-z')^2$ yields a non-vanishing contribution to the last integral, 
so that
$$\int d\sigma_{\mu\nu}(x)\int d\sigma_{\mu\nu}(x')=-\frac12\oint dz_\mu \oint dz_\mu'(-2zz')=
\left(\int_0^s d\tau{\,}\dot z_\mu
z_\nu\right)^2.$$}. Then, 
by virtue of an elementary Fourier transform
$\int_x {\rm e}^{-\mu|x|+ipx}=\frac{12\pi^2\mu}{(p^2+\mu^2)^{5/2}}$,
one has
\be
\label{vv}
\langle W(C)\rangle = N_c\int\left[\prod\limits_{\mu<\nu}^{} {\cal D}B_{\mu\nu}{\,}
{\rm e}^{-\frac{N_c}{\pi^2\mu G}\int_x B_{\mu\nu}(-\partial^2+\mu^2)^{5/2}B_{\mu\nu}}\right]{\,}
{\rm e}^{\frac{i}{2}\int_x B_{\mu\nu}\Sigma_{\mu\nu}}\equiv N_c\left<{\rm e}^{\frac{i}{2}\int_x B_{\mu\nu}\Sigma_{\mu\nu}}\right>_B,
\ee
where $\Sigma_{\mu\nu}\equiv\Sigma_{\mu\nu}(x;C)=\frac12\int_0^s d\tau{\,} (z_\mu\dot z_\nu-z_\nu\dot z_\mu)
\delta(x-z(\tau))$ and $\int_x\equiv\int d^4x$. The exponential of interest thus reads
$${\rm e}^{\frac{i}{2}\int_x B_{\mu\nu}\Sigma_{\mu\nu}}={\rm e}^{\frac{i}{4}\int_0^s d\tau{\,}B_{\mu\nu}(z){\,}
(z_\mu\dot z_\nu-z_\nu\dot z_\mu)}={\rm e}^{\frac{i}{2}\int_0^s d\tau{\,}B_{\mu\nu}(z)z_\mu\dot z_\nu},$$
where the antisymmetry of $B_{\mu\nu}$ has been used at the final step. One recognizes in this formula a
Wilson loop corresponding to the Abelian gauge field 
\be
\label{vv1}
A_\nu(z)=\frac12z_\mu B_{\mu\nu}(z).
\ee
The strength tensor of this field is $F_{\mu\nu}=\partial_\mu A_\nu-\partial_\nu A_\mu=B_{\mu\nu}+C_{\mu\nu}$,
where $C_{\mu\nu}(z)=\frac12z_\lambda(\partial_\mu B_{\lambda\nu}-\partial_\nu B_{\lambda\mu})$.
In particular, owing to just the Abelian Stokes' theorem, it is the strength tensor $F_{\mu\nu}$ which automatically 
appears in the quark spin term of the effective action, being recovered by the operator $\frac{\delta}{\delta\sigma_{\mu\nu}}$ in Eq.~(\ref{effA}).

Accordingly, the one-loop effective action~(\ref{effA}) takes the form
\be
\label{me}
\langle \Gamma[A_\mu^a]\rangle=2N_c\int_0^\infty\frac{ds}{s}{\,}\frac{{\rm e}^{-M^2s}}{(4\pi)^2}
\left<\int_x F_{\mu\nu}(x){\,}{\sf F}(\xi){\,}F_{\mu\nu}(x)\right>_B,
\ee
with the corresponding Abelian covariant derivative $D_\mu=\partial_\mu-iA_\mu$ entering the formfactor ${\sf F}(\xi)$.
In the spinor case at issue, this formfactor reads~\cite{1}
${\sf F}(\xi)=\frac{f(\xi)-1}{2\xi}-\frac14f(\xi)$, where $f(\xi)=\int_0^1 du{\,}{\rm e}^{u(1-u)\xi}$ and
$\xi=sD_\mu^2$. In what follows, we find convenient to 
identically represent the formfactor ${\sf F}(\xi)$ in the form
$${\sf F}(\xi)=\frac12\int_0^1 du\left[u(1-u)\int_0^1 d\alpha{\,}{\rm e}^{\alpha u(1-u)\xi}-
\frac12{\,}{\rm e}^{u(1-u)\xi}\right].$$
Following the method of Ref.~\cite{bc},
each of the two exponentials in the last expression can be represented as
$${\sf F}(\xi)=
\frac12\int_0^1 du\left[u(1-u)\int_0^1 d\alpha{\,}\frac{1}{[4\pi\alpha u(1-u)s]^2}\int_y {\rm e}^{-\frac{y^2}{4\alpha
u(1-u)s}+y_\mu D_\mu}-\right.$$
$$
\left.-\frac12{\,}\frac{1}{[4\pi u(1-u)s]^2}\int_y {\rm e}^{-\frac{y^2}{4u(1-u)s}+y_\mu D_\mu}
\right].$$
At the final step of the transformation, by performing an elementary $\alpha$-integration, we obtain for the 
formfactor ${\sf F}(\xi)$ a compact expression
\be
\label{11}
{\sf F}(\xi)=\frac{1}{2(4\pi s)^2}\int_0^1 du\int_y\left(\frac{4s}{y^2}-\frac{1}{2[u(1-u)]^2}\right)
{\rm e}^{-\frac{y^2}{4u(1-u)s}+y_\mu D_\mu}.
\ee
We now insert this expression into Eq.~(\ref{me}). Since the gauge field enters Eq.~(\ref{11}) only via 
the exponential ${\rm e}^{y_\mu D_\mu}$, we obtain for the $B$-average in Eq.~(\ref{me}):
\be
\label{22}
\left<\int_x F_{\mu\nu}(x){\,}{\rm e}^{y_\mu D_\mu}F_{\mu\nu}(x)\right>_B=V\langle F_{\mu\nu}(0)F_{\mu\nu}(y)
\rangle_B
=V\left[\langle B_{\mu\nu}(0)B_{\mu\nu}(y)\rangle_B+\langle B_{\mu\nu}(0)C_{\mu\nu}(y)\rangle_B\right].
\ee
To obtain the last equality in Eq.~(\ref{22}) we have used the fact that 
$C_{\mu\nu}(0)=0$~\footnote{Rigorously speaking, the
correlation functions $\langle B_{\mu\nu}(0)B_{\mu\nu}(y)\rangle_B$ and $\langle B_{\mu\nu}(0)C_{\mu\nu}(y)\rangle_B$
contain the phase factor $\exp\left[i\int_0^y du_\mu A_\mu(u)\right]$. 
However, the Taylor expansion of such a phase factor would yield 
correlation functions of more than two $B_{\mu\nu}$'s. On the other hand, the use of the formfactor ${\sf F}$ corresponds to accounting for only two $B_{\mu\nu}$'s. For this reason, we must approximate the said phase 
factor by unity.}. The correlation functions that enter Eq.~(\ref{22}) now read
\be
\label{33}
\langle B_{\mu\nu}(0)B_{\mu\nu}(y)\rangle_B=12\int {\cal D}B{\,}
{\rm e}^{-\frac{N_c}{\pi^2\mu G}\int_x B(-\partial^2+\mu^2)^{5/2}B}B(0)B(y)=
\frac{G}{2N_c}{\,}{\rm e}^{-\mu|y|}
\ee
and
\be
\label{uu}
\langle B_{\mu\nu}(0)C_{\mu\nu}(y)\rangle_B=\frac{G}{8N_c}{\,}(y_\lambda\partial_\lambda){\,}{\rm e}^{-\mu|y|}=
-\frac{G}{8N_c}{\,}\mu|y|{\,}{\rm e}^{-\mu|y|},
\ee
where we have taken into account that an antisymmetric tensor has 12 components.

The large-$M$ limit at issue corresponds to 
$\xi\ll 1$ and $\mu|y|\ll 1$, so that
$f(\xi)\to 1+\frac{\xi}{6}$, ${\sf F}(\xi)\to-\frac{1}{6}$, and only Eq.~(\ref{33}) 
contributes to Eq.~(\ref{22}), whereas Eq.~(\ref{uu}) does not. 
Recalling finally the definition of the quark condensate,
Eq.~(\ref{qq}), we recover Eq.~(\ref{hee}).

\begin{figure}
\psfrag{x}{\Large{$\lambda$}}
\psfrag{y}{\Large{$I(\lambda)$}}
\epsfig{file=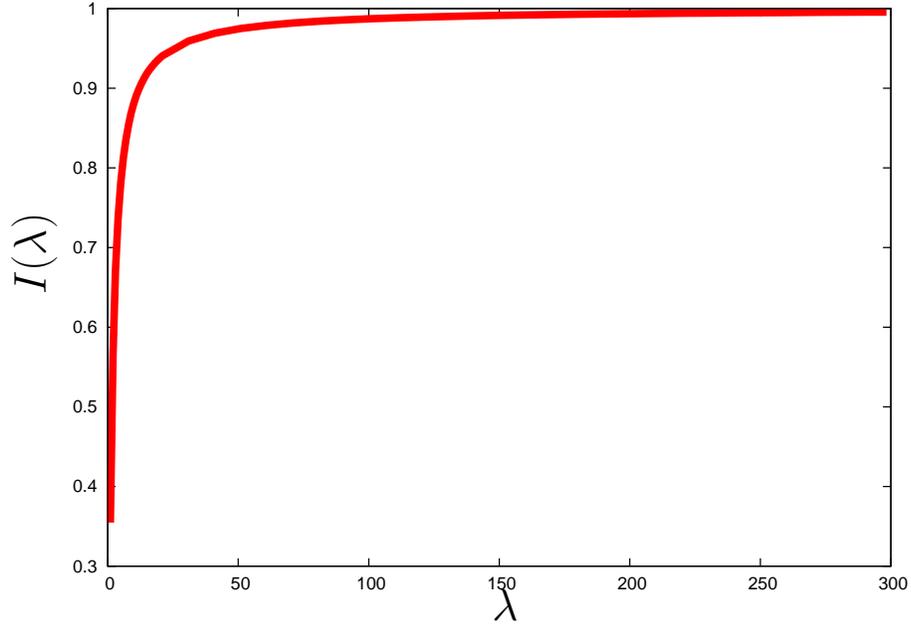, width=120mm}
\caption{The function $I(\lambda)$ in the range $\lambda\in[1,298]$, where $298\simeq\frac{M_t}{\mu}$ is the maximum value of 
$\lambda$, which corresponds to $T_g=0.34{\,}{\rm fm}$.}
\label{4}
\end{figure}

\begin{figure}
\psfrag{x}{\Large{$\lambda$}}
\psfrag{y}{\Large{$\frac{\langle\bar\psi\psi\rangle}{\langle\bar\psi\psi\rangle_{{\rm SVZ},f}}$}}
\epsfig{file=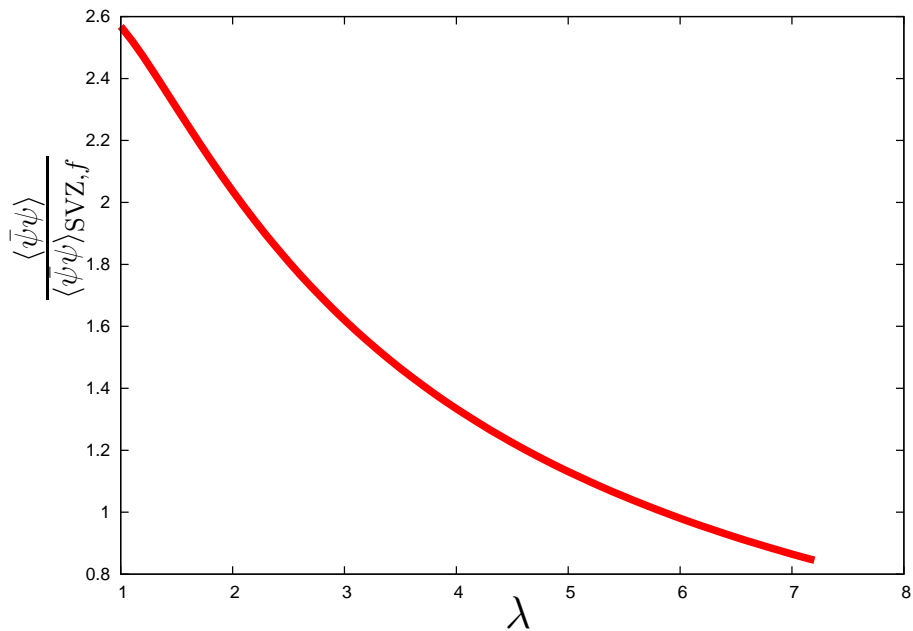, width=120mm}
\caption{Ratio (\ref{mr}) for $f=b$ and $T_g=0.34{\,}{\rm fm}$, in the range of masses $M\in[\mu,M_b]$.
The corresponding maximum value of $\lambda$ is $\frac{M_b}{\mu}\simeq7.24$.}
\label{br}
\end{figure}

We will now apply the same method of calculation of the effective action to a derivation of the 
quark condensate for the smaller values of $M$, down to $M=\mu$. Equations~(\ref{11})-(\ref{uu}) yield
$$\left<\int_x F_{\mu\nu}(x){\,}{\sf F}(\xi){\,}F_{\mu\nu}(x)\right>_B=$$
\be
\label{B1}
=\frac{VG}{4N_c(4\pi s)^2}
\int_0^1 du\int_y\left(\frac{4s}{y^2}-\frac{1}{2[u(1-u)]^2}\right)
{\rm e}^{-\frac{y^2}{4u(1-u)s}-\mu|y|}\left(1-\frac{\mu|y|}{4}\right),
\ee
where ``1'' in the last bracket stems from Eq.~(\ref{33}), while $\left(-\frac{\mu|y|}{4}\right)$ stems from 
Eq.~(\ref{uu}).
Accordingly, by using Eq.~(\ref{qq}), we can obtain for the quark condensate the following expression:
\bea
\langle\bar\psi\psi\rangle=
\frac{MG}{(4\pi)^4}\int_0^\infty ds{\,}\frac{{\rm e}^{-M^2s}}{s^2}
\int_0^1 du\int_y\left(\frac{4s}{y^2}-\frac{1}{2[u(1-u)]^2}\right)
{\rm e}^{-\frac{y^2}{4u(1-u)s}-\mu|y|}\left(1-\frac{\mu|y|}{4}\right).\nonumber
\eea
The $s$-integration in this formula can be performed exactly. Denoting 
$$z\equiv\mu|y|,~~~ \lambda\equiv\frac{M}{\mu},~~~ {\rm and}~~~ a\equiv\frac{\lambda}{\sqrt{u(1-u)}},$$
we arrive at the following intermediate result:
\be
\label{55}
\langle\bar\psi\psi\rangle=\frac{\lambda^2}{64\pi^2}{\,}\frac{G}{M}\int_0^1 du\int_0^\infty dz{\,}z{\,}
{\rm e}^{-z}
\left[4{\,}K_0(az)-\frac{az}{u(1-u)}{\,}K_1(az)\right]\left(1-\frac{z}{4}\right),
\ee
where $K_\nu$'s are the Macdonald functions.
The $z$-integration here 
can still be performed analytically, yielding 
\be
\label{hg44}
\frac{\langle\bar\psi\psi\rangle}{\langle\bar\psi\psi\rangle_{\rm SVZ}}\equiv I(\lambda),
\ee
where $\langle\bar\psi\psi\rangle_{\rm SVZ}$ is given by Eq.~(\ref{hee}), and 
$I(\lambda)$ stands for the following integral:
$$I(\lambda)=\frac{3\lambda^2}{4}\int_0^1\frac{du}{1-a^2}\left\{4+\left(\frac{a}{\lambda}\right)^2\cdot
\frac{2a^2+1}{1-a^2}+
\frac{3}{a^2-1}-\left(\frac{a}{2\lambda}\right)^2\cdot\frac{13a^2+2}{(a^2-1)^2}+\right.$$
$$\left.+
\frac{\arccos(1/a)}{(a^2-1)^{3/2}}\left[\frac{3a^4}{\lambda^2}-5a^2+2+\left(\frac{a}{2\lambda}\right)^2\cdot
\frac{3a^2(a^2+4)}{a^2-1}\right]\right\}.$$
For $\lambda\gg 1$, the leading large-$\lambda$ terms $4+\left(\frac{a}{\lambda}\right)^2\cdot
\frac{2a^2+1}{1-a^2}$ yield $I(\lambda)\to 1$.
For arbitrary $\lambda$'s, the remaining $u$-integration has been done numerically, with the result plotted 
in Fig.~\ref{4}. In particular, we obtain $I(1)\simeq0.36$, that is, a 64\%-decrease in the value of the 
quark condensate when $M\simeq\mu$.
In reality, only the values of $I(\lambda)$ corresponding to $M=M_c$, $M_b$, and $M_t$ are 
of physical significance. We use the standard quark masses $M_c\simeq1.3{\,}{\rm GeV}$, $M_b\simeq4.2{\,}{\rm GeV}$, and $M_t\simeq173{\,}{\rm GeV}$. The vacuum correlation length in full QCD with 
light flavors~\cite{gna}, $T_g\simeq0.34{\,}{\rm fm}$, corresponds to $\mu\simeq580{\,}{\rm MeV}$. This yields
\be
\label{n1}
I(M_c/\mu)\simeq 0.60,~~~ I(M_b/\mu)\simeq 0.84,~~~ I(M_t/\mu)\simeq 0.996~~~ {\rm in}~~ {\rm full}~~ {\rm QCD}.
\ee
For the alternative case of quenched QCD, that is, SU(3) pure Yang--Mills theory, 
the vacuum correlation length is~\cite{que} $T_g\simeq0.22{\,}{\rm fm}$, which corresponds to 
$\mu\simeq897{\,}{\rm MeV}$.
For this value of $\mu$, we have 
\be
\label{n2}
I(M_c/\mu)\simeq 0.47,~~~ I(M_b/\mu)\simeq 0.77,~~~ I(M_t/\mu)\simeq 0.993~~~ {\rm in}~~ {\rm quenched}~~ 
{\rm QCD}.
\ee
The sets of numbers~(\ref{n1}) and (\ref{n2}) illustrate the degree of accuracy of Eq.~(\ref{hee}) for various heavy flavors and various values of the vacuum correlation length $T_g$.
Since the case of heavy quarks considered here constitutes an intermediate case 
between QCD with light quarks and quenched QCD,
the genuine value of $I(M_f/\mu)$, for a given heavy flavor $f$, lies somewhere in between the two corresponding 
values of $I(M_f/\mu)$ listed in Eqs.~(\ref{n1}) and (\ref{n2}). In any case, we can conclude that Eq.~(\ref{hee})
is inapplicable to the $c$-quark, since it can develop up to 53\%-corrections [cf. $I(M_c/\mu)$ from Eq.~(\ref{n2})].
We notice that a qualitatively similar conclusion has been drawn in Ref.~\cite{shsi}, where the leading correction to Eq.~(\ref{hee}) has been evaluated through a non-perturbative gluon propagator in the Fock--Schwinger gauge.
Finally, setting in Eq.~(\ref{hee}) a certain heavy flavor $f$, and denoting 
$\langle\bar\psi\psi\rangle_{{\rm SVZ},f}\equiv-\frac{G}{48\pi^2 M_f}$,
we can write, instead of
Eq.~(\ref{hg44}),
\be
\label{mr}
\frac{\langle\bar\psi\psi\rangle}{\langle
\bar\psi\psi\rangle_{{\rm SVZ},f}}=\frac{M_f}{M}{\,}I(M/\mu).
\ee
For an illustration, we plot in Fig.~\ref{br} the ratio~(\ref{mr}) for the case of $f=b$
and $T_g=0.34{\,}{\rm fm}$, up to $M=M_b$.
In accordance with the intuitive expectations about the behavior of $\langle\bar\psi\psi\rangle$ with $M$, we observe a monotonic decrease of $|\langle\bar\psi\psi\rangle|$ with the increase of $M$.

\section{Accounting for the non-confining non-perturbative interactions}

In addition to the confining interactions of stochastic background Yang--Mills fields, 
which lead to the Wilson loop in the form of Eq.~(\ref{ww}), there also exist 
non-confining non-perturbative interactions of those fields. In this section, we demonstrate the
interesting phenomenon of complete independence of 
the quark condensate from such interactions, provided they exhibit exponential correlations.

To account for the non-confining non-perturbative interactions, one represents
the full two-point correlation function of gluonic field strengths in the form~\cite{ri, ds, ap} 
\bea
&&
\left<g^2F_{\mu\nu}^a(0)F_{\lambda\rho}^b(x)\right>=\nonumber\\
&&=\frac{G}{12}\cdot
\frac{\delta^{ab}}{N_c^2-1}\cdot
\bigl\{\kappa(\delta_{\mu\lambda}\delta_{\nu\rho}-\delta_{\mu\rho}\delta_{\nu\lambda})
+\frac{1-\kappa}{2}\left[\partial_\mu(x_\lambda\delta_{\nu\rho}-x_\rho\delta_{\nu\lambda})+
\partial_\nu(x_\rho\delta_{\mu\lambda}-x_\lambda\delta_{\mu\rho})\right]\bigr\}{\rm e}^{-\mu|x|}.\nonumber
\eea
Here, $\kappa\in[0,1]$ is some parameter, which defines the relative strength of the 
confining and the non-confining non-perturbative interactions. The lattice simulations in the SU(3) Yang--Mills 
theory yield the value of $\kappa=0.83\pm0.03$ (cf. Ref.~\cite{ri}), which means that the relative contribution of the non-confining non-perturbative interactions amounts to only 17\%. Expressing the Wilson loop via the correlation 
function
$\left<g^2F_{\mu\nu}^a(0)F_{\lambda\rho}^b(x)\right>$ through the non-Abelian Stokes' theorem and the cumulant expansion~\cite{ds}, and using the above parametrization of 
$\left<g^2F_{\mu\nu}^a(0)F_{\lambda\rho}^b(x)\right>$, 
one obtains the following generalization of Eq.~(\ref{ww}):
\bea
&&
\langle W(C)\rangle=N_c{\,}\exp\biggl\{-\frac{G}{96N_c}\times\nonumber\\
&&
\times\left[\kappa\int_\Sigma d\sigma_{\mu\nu}(x)
\int_\Sigma d\sigma_{\mu\nu}(x')+
\frac{1-\kappa}{\mu^2}\oint_C dx_\mu\oint_C dx_\mu'
(1+\mu|x-x'|)\right]{\rm e}^{-\mu|x-x'|}\biggr\}.
\label{dd}
\eea
The non-confining non-perturbative interactions produce in the Wilson loop a term with the double contour integral, which 
initially has the form (cf. Ref.~\cite{ap})
$\frac{1-\kappa}{2}\oint_C dx_\mu\oint_C dx_\mu'\int\limits_{(x-x')^2}^\infty d\tau{\,}
{\rm e}^{-\mu\sqrt{\tau}}$. The corresponding expression in Eq.~(\ref{dd}) resulted from the $\tau$-integration 
in this formula. 

Much as for the surface-dependent part of the Wilson loop, for the contour-dependent part we can also use 
some elementary Fourier transform, namely $\int_x (1+\mu|x|){\rm e}^{-\mu|x|+ipx}=\frac{60\pi^2\mu^3}{(p^2+\mu^2)^{7/2}}$,
to represent it as
$$\exp\left\{-\frac{(1-\kappa)G}{96N_c\mu^2}\int_{x,x'}j_\mu^x j_\mu^{x'}
(1+\mu|x-x'|){\rm e}^{-\mu|x-x'|}\right\}=
\int {\cal D}h_\mu{\,} {\rm e}^{-\frac{2N_c}{5\pi^2(1-\kappa)\mu G}
\int_x h_\mu(-\partial^2+\mu^2)^{7/2}h_\mu+i\int_x h_\mu j_\mu},$$
where $j_\mu^x\equiv j_\mu(x;C)=\oint_C dx_\mu(\tau)\delta(x-x(\tau))$. Further introducing a notation 
for the mean value
$\left<\cdots\right>_h=\int {\cal D}h_\mu{\,} {\rm e}^{-\frac{2N_c}{5\pi^2(1-\kappa)\mu G}
\int_x h_\mu(-\partial^2+\mu^2)^{7/2}h_\mu}(\cdots)$, we notice that 
the full Wilson loop~(\ref{dd}) can be written as a product of two averages:
$$\langle W(C)\rangle  
=N_c\left<{\rm e}^{\frac{i}{2}\int_x B_{\mu\nu}\Sigma_{\mu\nu}}\right>_B\cdot
\left<{\rm e}^{i\int_x h_\mu j_\mu}
\right>_h.$$ 
This equation generalizes Eq.~(\ref{vv}) to the case where the non-confining non-perturbative interactions are also taken into account.
Accordingly, the auxiliary Abelian gauge field~(\ref{vv1}) becomes now
$A_\nu(z)=\frac12z_\mu B_{\mu\nu}(z)+h_\nu(z)$. Its strength tensor reads $F_{\mu\nu}=B_{\mu\nu}+C_{\mu\nu}+
H_{\mu\nu}$, where $H_{\mu\nu}=\partial_\mu h_\nu-\partial_\nu h_\mu$. Furthermore, Eq.~(\ref{22}) also gets modified as 
$$\left<\int_x F_{\mu\nu}(x){\,}{\rm e}^{y_\mu D_\mu}F_{\mu\nu}(x)\right>_{B,h}=V\cdot \left[\langle B_{\mu\nu}(0)B_{\mu\nu}(y)\rangle_B+\langle B_{\mu\nu}(0)C_{\mu\nu}(y)\rangle_B+\langle H_{\mu\nu}(0)H_{\mu\nu}(y)\rangle_h\right],$$
where we have denoted $\langle\langle\cdots\rangle_B\rangle_h$ as just $\langle\cdots\rangle_{B,h}$.
The appearing additional correlation function 
$\langle H_{\mu\nu}(0)H_{\mu\nu}(y)\rangle_h$
can be readily calculated by means of the formula
$$\langle h_\mu(0)h_\nu(y)\rangle_h=\delta_{\mu\nu}{\,}\frac{(1-\kappa)G}{48N_c\mu^2}\cdot(1+\mu|y|)\cdot
{\rm e}^{-\mu|y|}.$$  
The result reads
$$\langle H_{\mu\nu}(0)H_{\mu\nu}(y)\rangle_h=\frac{(\kappa-1)G}{8N_c}\cdot(\mu|y|-4)\cdot{\rm e}^{-\mu|y|}.$$
Using now Eqs.~(\ref{33}) and (\ref{uu}), with $G$ replaced by $\kappa G$, 
we observe a remarkable mutual cancellation among all the 
$\kappa$-dependent contributions. Namely, we obtain
$$\frac{1}{V}\left<\int_x F_{\mu\nu}(x){\,}{\rm e}^{y_\mu D_\mu}F_{\mu\nu}(x)\right>_{B,h}=$$
$$=\frac{\kappa G}{2N_c}\left(1-\frac{\mu|y|}{4}\right){\rm e}^{-\mu|y|}+
\frac{(\kappa-1)G}{8N_c}\cdot(\mu|y|-4)\cdot{\rm e}^{-\mu|y|}=
\frac{G}{2N_c}{\rm e}^{-\mu|y|}
\left(1-\frac{\mu|y|}{4}\right).$$
Thus, Eq.~(\ref{B1}), with $\langle\cdots\rangle_B$ replaced by $\langle\cdots\rangle_{B,h}$, stays unchanged, and
so does the resulting quark condensate. 

The question whether the obtained cancellation  
among the $\kappa$-dependent contributions is specific for the above-considered exponential ansatz for the 
correlation function $\left<g^2F_{\mu\nu}^a(0)F_{\lambda\rho}^b(x)\right>$, or it holds equally well for other 
ans\"atze (such as e.g. the Gaussian one), requires a separate study, which lies beyond the scope of the present paper. We only notice that, even in the absence of such a cancellation,
the contribution of non-confining non-perturbative 
interactions is always suppressed, in comparison with the contribution of confining interactions, by a relative factor
of $\frac{1-\kappa}{\kappa}\simeq0.2$.

\section{Summary}

The aim of the present paper was to find a relation between the quark and the gluon condensates, 
which would yield, for various heavy flavors, corrections to the known Eq.~(\ref{hee}). 
The corrections thus obtained, given by Eqs.~(\ref{n1}) and (\ref{n2}), show that Eq.~(\ref{hee}) applies with a good accuracy only to the $t$-quark. Rather, for the $b$-quark, the corrections are $\sim 20\%$, while for the $c$-quark
they can be as large as $\sim 50\%$, thereby making Eq.~(\ref{hee}) inapplicable to the $c$- and the $s$-quarks. 
Also, as one can see from Fig.~1, when the continuously varied current quark mass $M$ reaches the value
of the inverse vacuum correlation length $\mu$, the absolute value of the quark condensate decreases by 64\% compared 
to the value provided by Eq.~(\ref{hee}). 

We have used in our calculations the most general
ansatz for the Wilson loop, which is provided by the stochastic vacuum model and accounts for the 
confining and non-perturbative non-confining interactions of the stochastic gluonic fields. The corresponding 
two-point surface-surface and contour-contour self-interactions 
of the Wilson loop can be represented as being mediated by an auxiliary 
Abelian gauge field with the Gaussian action. In particular, for the most simple, exponential, parametrization 
of the two-point correlation function of gluonic field strengths, we have found an interesting phenomenon of 
a complete independence of the heavy-quark condensate from the non-confining non-perturbative interactions of the stochastic gluonic fields.

In conclusion, we have started our analysis from the heavy-quark limit, where chiral symmetry 
is explicitly broken by a large current quark mass. The advantage of working in this limit is that one 
avoids possible uncertainties related to the particular form of the 
field-strength correlation function. Indeed, owing 
to the constancy of the gauge field inside the heavy-quark trajectory, Eq.~(\ref{hee}) in the $t$-quark case 
turns out to be almost exact. We emphasize that even in the heavy-quark limit we still have a relation connecting 
the quark condensate $\langle\bar\psi\psi\rangle$ with the gluon condensate
$\langle (gF_{\mu\nu}^a)^2\rangle$.
We have not proceeded to the current quark masses smaller than $\mu$ (cf. Fig.~1), which is the case of $s$-, $d$-, and $u$-quarks. The reason is that, for such light quarks, the effect of spontaneous breaking 
of chiral symmetry starts to play an 
important role, resulting in the appearance of a significant self-energy contribution to
the dynamical constituent quark mass. Thus, since such a self-energy contribution cannot be consistently calculated 
within the adopted world-line formalism, we have to restrict our analysis to the case of heavy quarks,
for which this contribution can be safely disregarded compared to the current quark mass.
However, even in the heavy-quark case provided by the 
$b$- and $c$-quarks, we have found substantial corrections to Eq.~(\ref{hee}).
The way in which Eq.~(\ref{hee}) along with these corrections 
goes over into Eq.~(\ref{kj}) for light quarks can be 
the subject of a separate study.

\begin{acknowledgments}

\noindent
One of us (D.A.) is grateful for the stimulating discussions to O.~Nachtmann and M.G.~Schmidt. 
The work of D.A. was supported by the Portuguese Foundation for Science and Technology
(FCT, program Ci\^encia-2008) and by 
the Center for Physics of Fundamental Interactions (CFIF) at Instituto Superior
T\'ecnico (IST), Lisbon. 
\end{acknowledgments}


\begin{thebibliography}{999} 
 
\bibitem{ne}
N.~Brambilla and A.~Vairo,
Phys.\ Lett.\ B {\bf 407}, 167 (1997);
P.~Bicudo, N.~Brambilla, J.E.F.T.~Ribeiro and A.~Vairo,
Phys.\ Lett.\  B {\bf 442}, 349 (1998).


\bibitem{chr}
Z.~Bern and D.~A.~Kosower, Phys.\ Rev.\ Lett.\  {\bf 66}, 1669 (1991);
Nucl.\ Phys.\  B {\bf 379}, 451 (1992);
M.~J.~Strassler, Nucl.\ Phys.\  B {\bf 385}, 145 (1992);
for reviews see: M.~Reuter, M.~G.~Schmidt and C.~Schubert, Annals Phys.\  {\bf 259}, 313 (1997);
C.~Schubert, Phys.\ Rept.\  {\bf 355}, 73 (2001).


\bibitem{2}
D.~Antonov and J.~E.~F.~T.~Ribeiro,
Phys.\ Rev.\  D {\bf 81}, 054027 (2010).


\bibitem{3}
M.~A.~Shifman, A.~I.~Vainshtein and V.~I.~Zakharov,
Nucl.\ Phys.\ B {\bf 147}, 385 (1979);
for a review see: S.~Narison, {\it QCD spectral sum rules}, World Scientific, 1989.


\bibitem{1}
M.~G.~Schmidt and C.~Schubert,
Phys.\ Lett.\  B {\bf 318}, 438 (1993); for the bosonic case see: A.~O.~Barvinsky and G.~A.~Vilkovisky,
Nucl.\ Phys.\ B {\bf 333}, 471 (1990).


\bibitem{ri}
E.~Meggiolaro, Phys.\ Lett.\ B {\bf 451}, 414 (1999).


\bibitem{mm}
Yu.~M.~Makeenko and A.~A.~Migdal, Phys.\ Lett.\  B {\bf 88}, 135 (1979);
Nucl.\ Phys.\  B {\bf 188}, 269 (1981).


\bibitem{ds}
H.~G.~Dosch and Yu.~A.~Simonov, Phys.\ Lett.\ B {\bf 205}, 339 (1988).


\bibitem{bc}
V.~I.~Shevchenko, JHEP {\bf 03}, 082 (2006).

\bibitem{gna}
M.~D'Elia, A.~Di Giacomo and E.~Meggiolaro,
Phys.\ Lett.\  B {\bf 408}, 315 (1997).

\bibitem{que}
A.~Di Giacomo and H.~Panagopoulos,
Phys.\ Lett.\ B {\bf 285}, 133 (1992);
A.~Di Giacomo, E.~Meggiolaro and H.~Panagopoulos,
Nucl.\ Phys.\ B {\bf 483}, 371 (1997).


\bibitem{shsi}
V.~Shevchenko and Yu.~Simonov,
Phys.\ Rev.\ D {\bf 65}, 074029 (2002).


\bibitem{ap}
D.~Antonov,
Annals Phys.\  {\bf 325}, 1304 (2010).
%%CITATION = APNYA,325,1304;%%





%\bibitem{st1}
%S.~Narison, Phys.\ Lett.\ B {\bf 387}, 162 (1996).


%\bibitem{st}
%F.~E.~Close, {\it An Introduction to quarks and partons}, Academic Press, 1979.

%\bibitem{gl2}
%N.~Brambilla, A.~Pineda, J.~Soto and A.~Vairo,
%Nucl.\ Phys.\ B {\bf 566}, 275 (2000);
%Yu.~A.~Simonov, Nucl.\ Phys.\ B {\bf 592}, 350 (2001);
%D.~Antonov, Phys.\ Lett.\ B {\bf 696}, 214 (2011).


\end{thebibliography}
\end{document}